\documentclass[conference,a4paper]{IEEEtran}
\IEEEoverridecommandlockouts
\usepackage[left=1.59cm,right=1.59cm,top=1.9cm,bottom=4.5cm]{geometry}
\usepackage{cite}
\usepackage{amsmath,amssymb,amsfonts}
\usepackage{algorithmic}
\usepackage{svg}
\usepackage{graphicx}
\usepackage{subcaption}
\usepackage{textcomp}
\usepackage{xcolor}
\usepackage{tabularx}
\usepackage{lipsum}
\usepackage{float}
\usepackage{stmaryrd}
\usepackage{flushend}

\def\BibTeX{{\rm B\kern-.05em{\sc i\kern-.025em b}\kern-.08em
    T\kern-.1667em\lower.7ex\hbox{E}\kern-.125emX}}
\usepackage{hyphenat} 

\newcommand{\imaginary}{{i\mkern1mu}}

\newcommand{\argmax}{\mathop{\mathrm{argmax}}\limits}

\newcommand{\IEEEcompact}{
\fontdimen3\font=0.2ex
\fontdimen2\font=0.5ex
}
\begin{document}

\title{Quasisynchronous LoRa  for LEO Nanosatellite Communications\\
}
\author{
\IEEEauthorblockN{
  	    ${}^{*}$Hasan Gökhan Uysal
        ${}^{\star}$Ferkan Yilmaz
  	    ${}^{\dag}$Hakan Ali \c{C}{\i}rpan
        ${}^{c}$O\u{g}uz Kucur
        ${}^{\ddag}$H\"{u}seyin Arslan}\\[-3mm]
  	\IEEEauthorblockA{
        ${}^{*\star\dag}${Istanbul Technical University, Electronics and Communication Engineering}, Istanbul, Turkey\\ 
  	    ${}^{\ddag}${Istanbul Medipol University, Electrical and Electronics Engineering}, Istanbul, Turkey\\ 
        ${}^{c}${Gebze Technical University, Electronics Engineering}, Gebze, Kocaeli\\ 
        ${}^{*}${Aselsan Inc.}, Ankara, Turkey\\[2mm]
 		    ${}^{*}$\texttt{gokhanuysal@aselsan.com.tr}, 
            \{\!
                ${}^{\star}$\texttt{yilmazf},                
                ${}^{\dag}$\texttt{cirpanh}
            \!\}\texttt{@itu.edu.tr},\\
            ${}^{c}$\texttt{okucur@gtu.edu.tr},
            ${}^{\ddag}$\texttt{huseyinarslan@medipol.edu.tr}
  	}
\vspace{-6mm}
}
   
\maketitle
\IEEEcompact
\begin{abstract}

Perfect synchronization in LoRa communications between Low Earth Orbit (LEO) satellites and ground base stations is still challenging, despite the potential use of atomic clocks in LEO satellites, which offer high precision. Even by incorporating atomic clocks in LEO satellites, their inherent precision can be leveraged to enhance the overall synchronization process, perfect synchronization is infeasible due to a combination of factors such as signal propagation delay, Doppler effects, clock drift and atmospheric effects. These challenges require the development of advanced synchronization techniques and algorithms to mitigate their effects and ensure reliable communication from\,/\,to LEO satellites. However, maintaining acceptable levels of synchronization rather than striving for perfection, quasisynchronous (QS) communication can be adopted which maintains communication reliability, improves resource utilization, reduces power consumption, and ensures scalability as more devices join the communication. Overall, QS communication offers a practical, adaptive, and robust solution that enables LEO satellite communications to support the growing demands of IoT applications and global connectivity. In our investigation, we explore different chip waveforms such as rectangular and raised cosine. Furthermore, for the first time, we study the Symbol Error Rate (SER) performance of QS LoRa communication, for different spreading factors (SF), over Additive White Gaussian Noise (AWGN) channels.

\end{abstract}


\section{Introduction}
\begin{figure}[tp]
    \centering
    \includegraphics[width=0.8\linewidth]{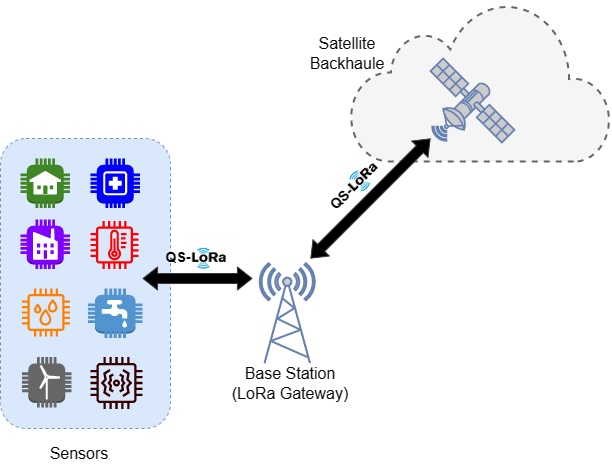}
    \caption{Quasisynchronous LoRa Scenarios}  
    \label{fig:1}
    \vspace{-5mm}
\end{figure}

In recent years, Low Earth Orbit (LEO) satellite communication has attracted a myriad of interest due to its ability to provide both high-performance communication and widespread location-based services around the world \cite{b1,b2,b3,b4,b5,b6}. Long-range technology (LoRa) is becoming one of the promising candidates for use in LEO systems due to its low power consumption and long range, making it suitable for a wide range of Internet of Things (IoT) applications. Linear Frequency Modulated (LFM) or chirp signals are frequently employed in radar transmissions, and more recently, they have been utilized in Low Power Wide Area Networks (LPWANs) for IoT applications. LoRa, the physical layer of the LPWAN system, conveys information through frequency shifts, with the initial frequency shift of the symbol serving as the information-carrying element \cite{b8,b9,b10,b12}. The overall system of LoRa transmission used in LEO satellite communications is illustrated in Fig. \ref{fig:1}.
However, achieving perfect synchronization in LoRa communications from ground stations to LEO satellites remains a challenge when taking into account the factors that cause this inability, such as signal propagation delay, Doppler effects, clock drift, and atmospheric effects. Particularly, one of the challenges for perfect synchronization comes from signal propagation delay. Even though LEO satellites are lower in altitude than GEO satellites, radio signals require a measurable amount of time to reach LEO satellites. This time delay varies depending on the relative positions of the LEO satellite and ground stations, making it difficult to achieve consistent synchronization. Since LEO satellites are constantly moving in their orbits, the distance from each ground station is constantly changing, leading to dynamic changes in signal propagation time\cite{b5}. Further, the Doppler effect contributes to the problem of synchronization with LEO satellite communications. Radio signals experience frequency shifts as satellites move in relation to the Earth. Accordingly, the apparent frequency of the signal received by an LEO satellite and the frequency of the signal transmitted from a ground station is slightly different, which can cause significant synchronization errors and ultimately result in communication failure\cite{b6}. Another factor affecting synchronization is clock drift. However, the synchronization is subjected to time drift, resulting in small differences in perceived time. These small errors can accumulate over time, leading to serious synchronization problems. Temperature fluctuations, mechanical stresses, and material aging can increase clock drift, making it difficult to maintain perfect synchronization\cite{b7}. Finally, atmospheric effects also affect the synchronization of communications from ground stations to LEO satellites. Factors such as ionospheric and tropospheric delays, scintillation, and absorption can cause signal degradation and delays that affect the accuracy and reliability of the communication link. These atmospheric conditions can vary with respect to time and location, making perfect synchronization even more difficult\cite{b2}. Accordingly, signal propagation delay, Doppler effects, clock drift, and atmospheric effects cause perfect synchronization to be infeasible in LoRa communications between LEO satellites and ground base stations. Therefore, advanced synchronization techniques and algorithms are needed for reliable communication in LEO satellite communication. Quasisynchronous communication plays a crucial role in addressing the challenges associated with achieving perfect synchronization in LoRa communication between LEO satellites and ground base stations. By focusing on maintaining acceptable levels of synchronization rather than striving for perfection, quasisynchronous communication enhances communication reliability, improves resource utilization, reduces power consumption, and ensures scalability as more devices join the network. Furthermore, this approach provides resilience to interference and varying atmospheric conditions while simplifying network management. Overall, quasisynchronous communication offers a practical, adaptive, and robust solution that enables LEO satellite communications to support the growing demands of IoT applications and global connectivity. In this context, our contributions can be summarized as follows:

\begin{itemize}
    \item We propose that in LoRa communications, when perfect synchronization is infeasible, adopting quasisynchronous transmission offers benefits such as improved time synchronization accuracy and increased resilience against performance loss, particularly in environments with varying distances and propagation conditions. To demonstrate how these advantages contribute to a more reliable and efficient communication system, we conducted system performance simulations. 
    \item  Pulse shaping in quasisynchronous LoRa communications offers several benefits. It enhances spectral efficiency by controlling bandwidth and reducing spectral leakage and increases power efficiency and improves coexistence with other systems. We investigate the performance of quasisynchronous LoRa communications for well-known rectangular and raised cosine chip waveforms.    
\end{itemize}

The rest of the paper is organized as follows. In Section II, the system model of the quasisynchronous LoRa communication is presented including the transmitter design, the receiver design, the quasisynchronous analysis with pulse shaping and the recovery of LoRa samples in AWGN channel. Simulation results are provided in Section III. In the last section, the conclusions are drawn.

\begin{figure*}
    \centering
    \includegraphics[width=0.6\textwidth]{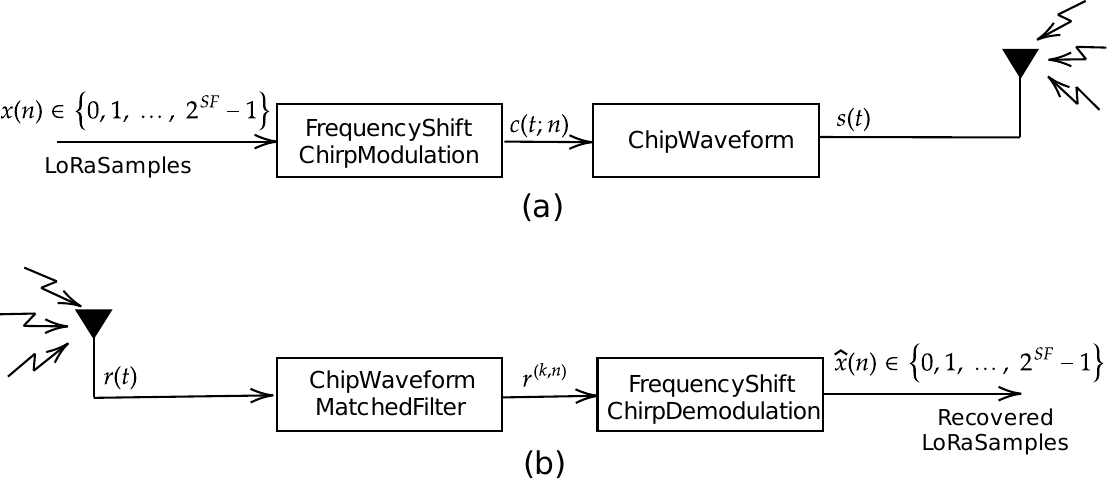}
    \caption{Quasisynchronous LoRa transmitter and receiver structures: (a) Transmitter, (b) Receiver}
    \label{fig:2}
\end{figure*}

\section{Quasisynchronous LoRa}
In LoRa, the information is carried by frequency shifts. In other words, the information carrying element is by frequency shifts at the beginning of symbols. Then, the chirp behaves like a carrier. Therefore, we can describe LoRa as a Frequency Shift Chirp Modulation (FSCM)\cite{b8,b9,b10}. Unlike the existing literature, the quasi-simultaneous LoRa transmitter and receiver structure that we propose and investigate in our paper is shown in Fig. \ref{fig:2}. Assuming the transmission bandwidth $B$ implies that each sample is transmitted through a channel at intervals of $T=1/B$. As such, let us denote the LoRa sample at time index $nT$, ($n\in\mathbb{Z}$) by $x(n)$ which is an integer number, taking values of $\{0,1,2,\ldots,2^{SF}-1\}$, formed using a word $w(n)$ of $\mathrm{SF}$ binary bits, that is 
\begin{equation}\label{Eq:LoRaSample}
    x(n)=\sum_{i=0}^{SF-1}2^{i}w(n;i),
\end{equation}
where $w(n;i)$ denotes the $i$-th bit of the word $w(n)$ of spreading factor ($\mathrm{SF}$) binary bits and $x(n)$ denotes the decimal conversion of te word $w(n)$. Hence, the LoRa symbol, to be transmitted at time index $nT$, is written in terms of the chip waveform as 
\begin{equation}\label{Eq:LoRaSymbol}
\!\!c(t;n)=\sqrt{P}\sum_{k=0}^{2^\mathrm{SF}-1}\!\!
            \Phi(x(n);k)
            \Psi_{T_c}(t-nT-kT_c),\!\!
\end{equation}
defined over $(n-1)T\leq{t}<nT$, where $P$ denotes the power of LoRa symbols, and $\Phi(x(n);k)\in\mathbb{C}$ denotes LoRa complex envelope given by   
\begin{equation}\label{Eq:phi}
\Phi(x(n);k)=\frac{1}{\sqrt{2^\mathrm{SF}}} 
        \exp\Bigl(\imaginary{2}\pi{k}\frac{(x(n)+k)_{2^\mathrm{SF}}}{2^\mathrm{SF}}\Bigr),
\end{equation}
where $\imaginary\equiv\sqrt{-1}$ denotes the imaginary number, and $(a)_b$ denotes the modulus function of $a$ with respect to $b$. Furthermore, $\Psi_{T_c}(t)$ represents any shape of chip waveform having unit energy and duration of $T_c\!=\!T/2^\mathrm{SF}$ seconds\cite{b11}. In this work, we obtain the quasisynchronous performance expressions for any chip-limited chip waveform but use rectangular and raised-cosine chip waveforms for simulation and numerical results; that is 
\begin{align}
\label{Eq:rect}
\!\!\!\!\!\text{Rectangular:}~~~\Psi_{T_c}(t)&=\sqrt{\frac{1}{T_c}}\prod_{T_c}(t),\\
\label{Eq:rcs}
\!\!\!\text{Raised-cosine:}\Psi_{T_c}(t)&=\sqrt{\frac{2}{3T_c}}[1-\cos\left({2\pi{t}}/{T_c}\right)]\prod_{T_c}(t),\!\!\!\!
\end{align}
where $\prod_{T_c}(t)$ denotes the unit amplitude waveform of duration $T_c$ seconds (i.e., $\prod_{T_c}(t)\!=\!1$ if $0\!\leq\!{t}\!<\!T_c$ otherwise $\prod_{T_c}(t)\!=\!0$). Thence, the total transmitted signal $s(t)$ can be written as  
\begin{equation} \label{Eq:tx}
\begin{split}
    s(t)&=\sum_{n}c(t;n)\\[-3mm]
    &=\sqrt{P}\sum_{n}\sum_{k=0}^{2^\mathrm{SF}-1}\!\!
            \Phi(x(n);k) \Psi_{T_c}(t-nT-kT_c).
\end{split}
\end{equation}
Upon using total transmitted signal, the total LoRa signal at the receiver can be written in terms of the chip waveform as
\begin{equation}
\label{Eq:received}
    r(t)=\sqrt{P}\sum_{n}\sum_{k=0}^{2^\mathrm{SF}-1}\!\!
            \Phi(x(n);k)
            \Psi_{T_c}(t-nT-kT_c) + n(t),
\end{equation}
where $n(t)$ is the AWGN with zero mean and the variance of $N_0$.

Considering the time synchronization between the LoRa receiver and the total received signal in \eqref{Eq:received}, it is important to note that the transmissions from the base station to LEO satellites and from various devices (such as mobile users, IoT devices, sensors, and factories) to the base station are not typically synchronous. This can be either quasisynchronous or asynchronous, resulting from varying distances to base stations and from base stations to LEO satellites. As a result, there exists a maximum time-synchronization error between transmitter and receiver. In asynchronous communication, the time-synchronization error is limited to one symbol duration, whereas in quasisynchronous communication it is restricted to a small fraction of one chip duration. Therefore, the time-synchronization error in quasisynchronous communication can be expressed as $\tau=\Delta{T_c}$, where $\Delta$ is the normalized time-synchronization error ($-\Delta_s/2\leq\Delta\leq\Delta_s/2$), where $\Delta_s$ denotes the maximum normalized time-synchronization error at the receiver such that $0\leq\Delta_s\leq{1}$. In accordance, the received signal $r(t)$ is matched-filtered with the chip-limited chip waveform $\Psi_{T_c}(t)$ and subsequently sampled at time instants $nT+kT_c$, ($n\in\mathbb{Z}$ and $k\in\{0,1,\ldots,2^\mathrm{SF}-1\}$). Accordingly, we write chip samples recovered by the receiver as;
\begin{equation}\label{Eq:matchfilter1}
r^{(k,n)}=\int_{nT+kT_c}^{nT+(k+1)T_c}\!\!\!\!r(t)\Psi_{T_c}(t-nT-kT_c-\Delta{T}_c)dt.
\end{equation}
After performing some algebraic manipulations, we can obtain \eqref{Eq:matchfilter1}, referring to Fig. \ref{fig:2}, in terms of partial correlation of chip waveform, partial cross-correlation of Lora symbols, and AWGN noise as follows
\begin{multline}\label{Eq:ChipFiltered}
r^{(k,n)}=\sqrt{P}\,\Phi(x(n);k)R_\Psi(\Delta)\\
            +\sqrt{P}\,\Phi(x(\hat{n});\hat{k})\hat{R}_\Psi(\Delta)
            +\sqrt{N_0T_c}\eta,
\end{multline}
with overlapped chip index $\hat{k}=k+\llbracket{\Delta}\rrbracket$ and overlapped symbol index $\hat{n}=n+\llbracket{k+\Delta}\rrbracket_{2^\mathrm{SF}}$, where $\llbracket{x}\rrbracket_{y}$ denotes the expanded sign function\cite[eq. (13)]{b11}, and $\llbracket{x}\rrbracket\equiv\llbracket{x}\rrbracket_{0}$ is the well known sign function. Moreover, $\eta$ denotes complex AWGN with zero mean and unit variance. $R_\Psi(\Delta)$ and $\hat{R}_\Psi(\Delta)$ are the partial auto-correlation functions of overlapping chips and overlapped chips\cite[Eqs. (14) and (15)]{b11}, respectively. Upon considering the intervals of LoRa symbols, and under the assumption of that the $m$th LoRa sample has been transmitted, the LoRa Symbol at time instant $nT$, denoted by $\hat{c}(t;n,m)$, can be recovered from LoRa samples as follows
\begin{equation}\label{Eq:LoRaFiltered}
    \hat{c}(t;n,m)=\sum_{k=0}^{2^\mathrm{SF}-1}r^{(k,n)}\Phi^{*}(m;k),
\end{equation}
with $m\!\in\!\{0,1,\ldots,2^\mathrm{SF}\!-\!1\}$, where the superscript $*$ denotes the complex conjugation. It is worthy noticing that we have $\sum_{k=0}^{2^\mathrm{SF}-1}\Phi(\hat{m};k)\Phi^{*}(m;k)=\delta_{\hat{m},m}$, where $\delta_{\cdot,\cdot}$ denotes Kronecker Delta function given by  
\begin{equation}\label{Eq:KrDelta}
\textstyle \delta_{\hat{m},m} =
    \begin{cases}
            1, &         \text{if } \hat{m}=m,\\
            0, &         \text{if } \hat{m}\neq m.
    \end{cases}
\end{equation}
Then, we reduce \eqref{Eq:LoRaFiltered} to 
\begin{align}\label{Eq:ChipFilteredExtended}
\nonumber
\hat{c}(t;n&,m)
=\sqrt{P}\,R_\Psi(\Delta)\delta_{x(n),m}
\sqrt{P}\,\hat{R}_\Psi(\Delta)R_{x(n),m}(\llbracket{\Delta}\rrbracket)\\
&~+\sqrt{P}\,\hat{R}_\Psi(\Delta)\hat{R}_{x(n+\llbracket{\Delta}\rrbracket),m}(\llbracket{\Delta}\rrbracket)+\sqrt{N_0T}\eta,
\end{align}
where $R_{\hat{m},m}(\ell)$ and $\hat{R}_{\hat{m},m}(\ell)$ are the partial cross-correlation functions between LoRa symbols to overlapping symbols (symbols with the same time index $n$) and overlapped adjacent LoRa symbols (symbols with the time index $n\pm{1}$)\cite[eq. (18) and (19)]{b11}. Finally, the recovered LoRa sample at the receiver can be written as 
\begin{equation}\label{Eq:decodedsample}
    \hat{x}(n)=\argmax_{m\in\{0,1,\ldots,2^\mathrm{SF}-1\}}\left|\hat{c}(t;n,m)\right|.
\end{equation}

\begin{figure*}[t]
    \centering
    \begin{subfigure}{0.45\textwidth}
        \includegraphics[width=\linewidth]{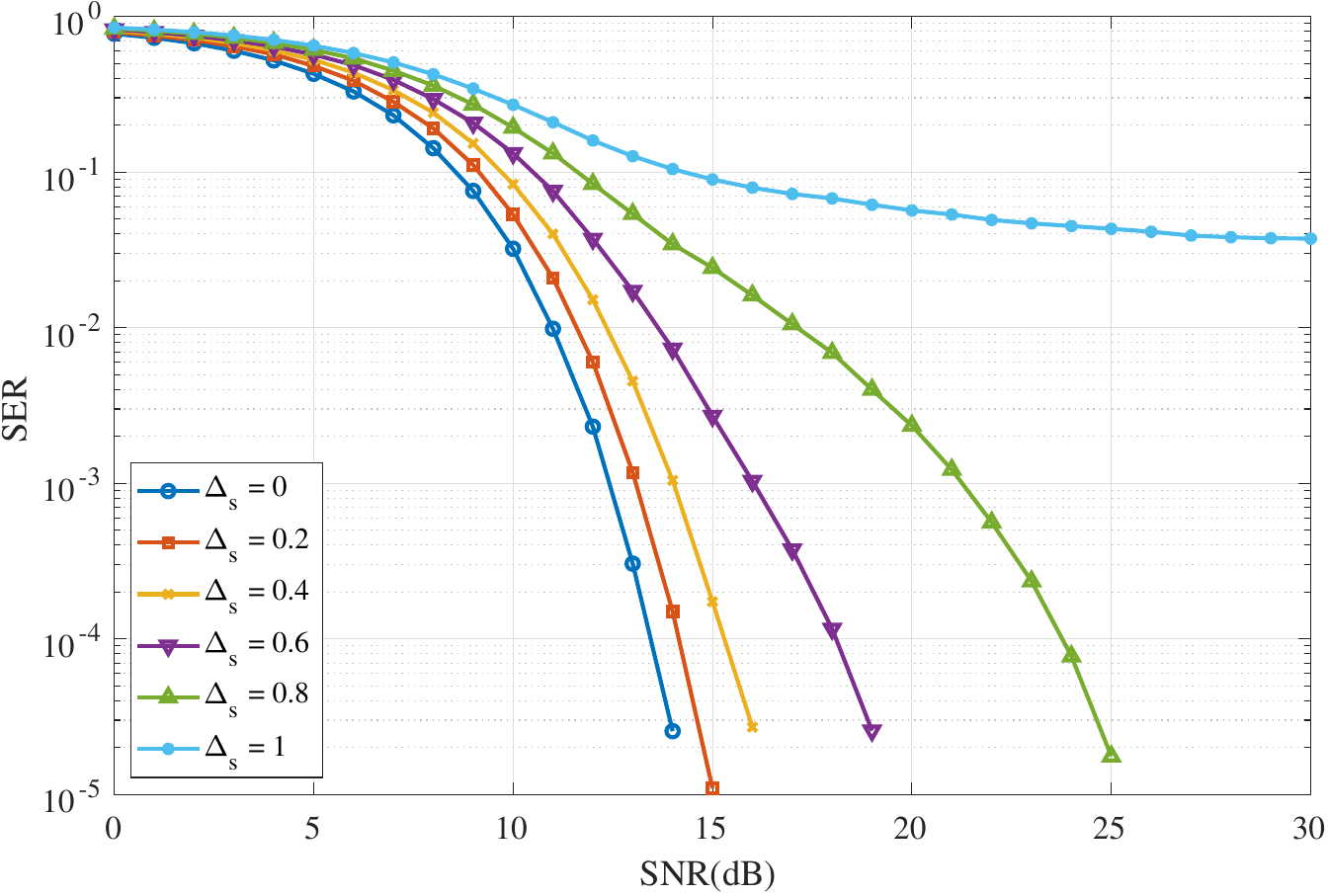}
        \caption{$\mathrm{SF}={4}$ and rectangular chip waveform}
        \label{fig:3}
    \end{subfigure}
    \hfill
    \begin{subfigure}{0.45\textwidth}
        \includegraphics[width=\linewidth]{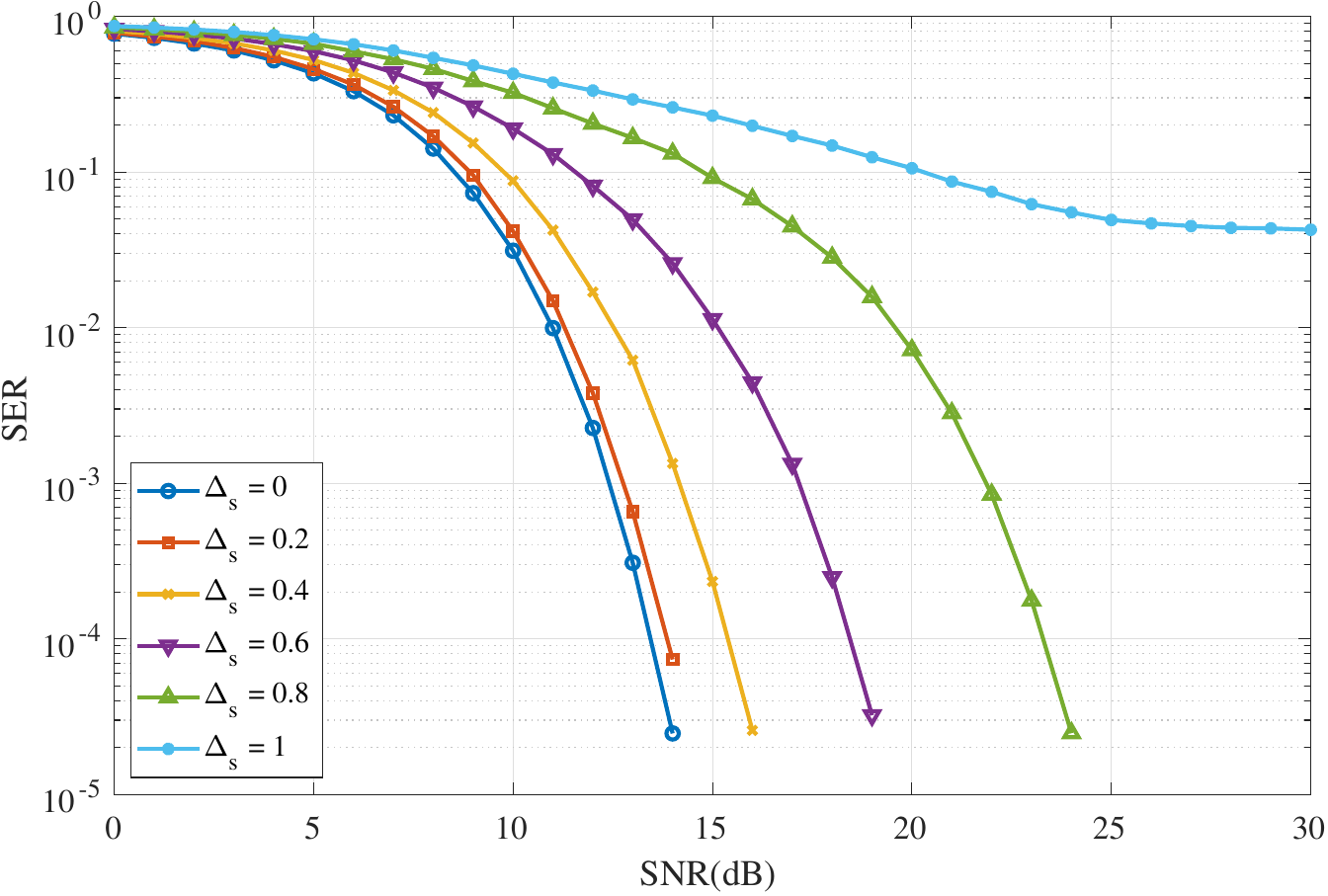}
        \caption{$\mathrm{SF}={4}$ and raised-cosine chip waveform}
        \label{fig:4}
    \end{subfigure}

    \vspace{6pt} 
     
    \begin{subfigure}{0.45\textwidth}
        \includegraphics[width=\linewidth]{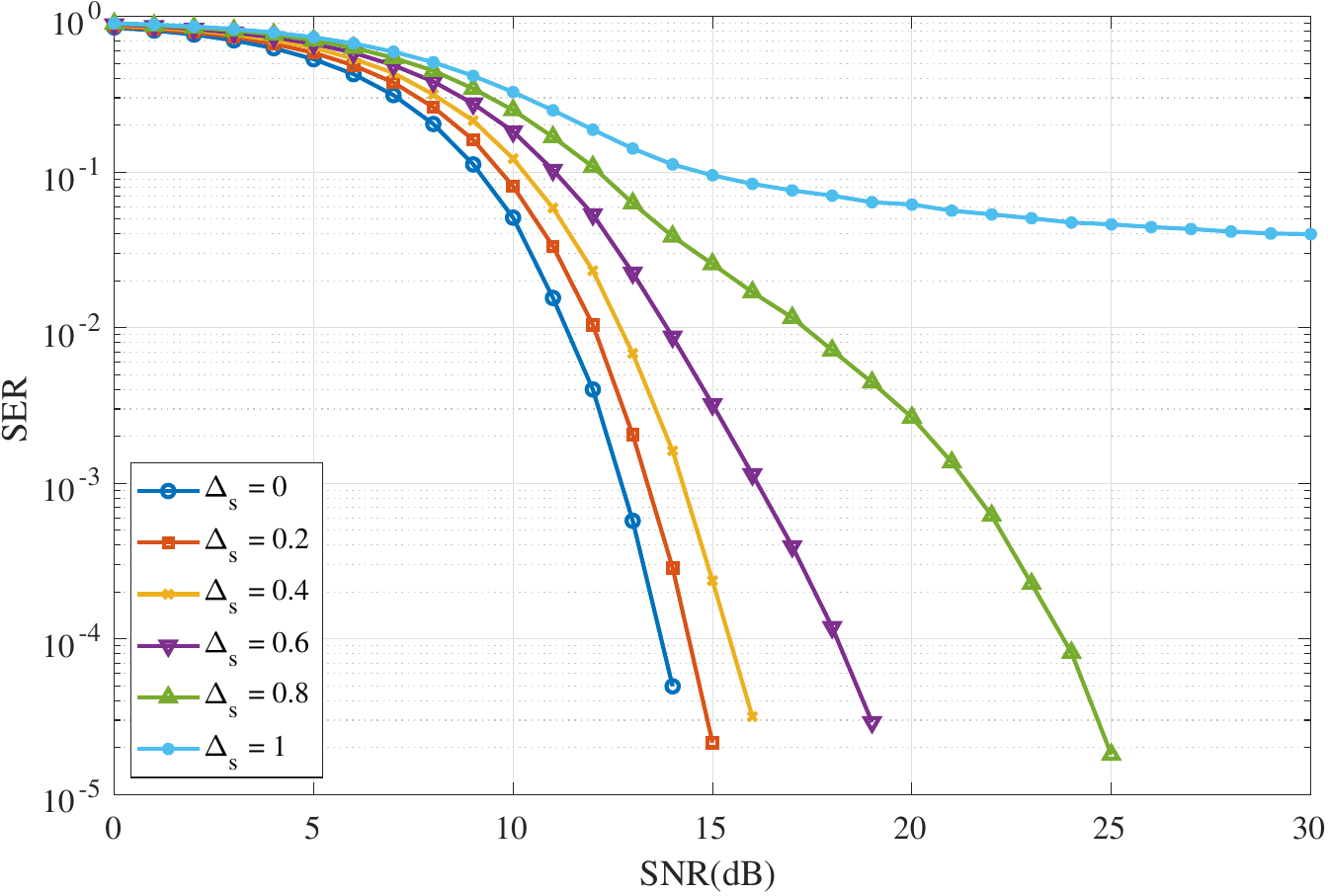}
        \caption{$\mathrm{SF}={5}$ and rectangular chip waveform}
        \label{fig:5}
    \end{subfigure}
    \hfill
    \begin{subfigure}{0.45\textwidth}
        \includegraphics[width=\linewidth]{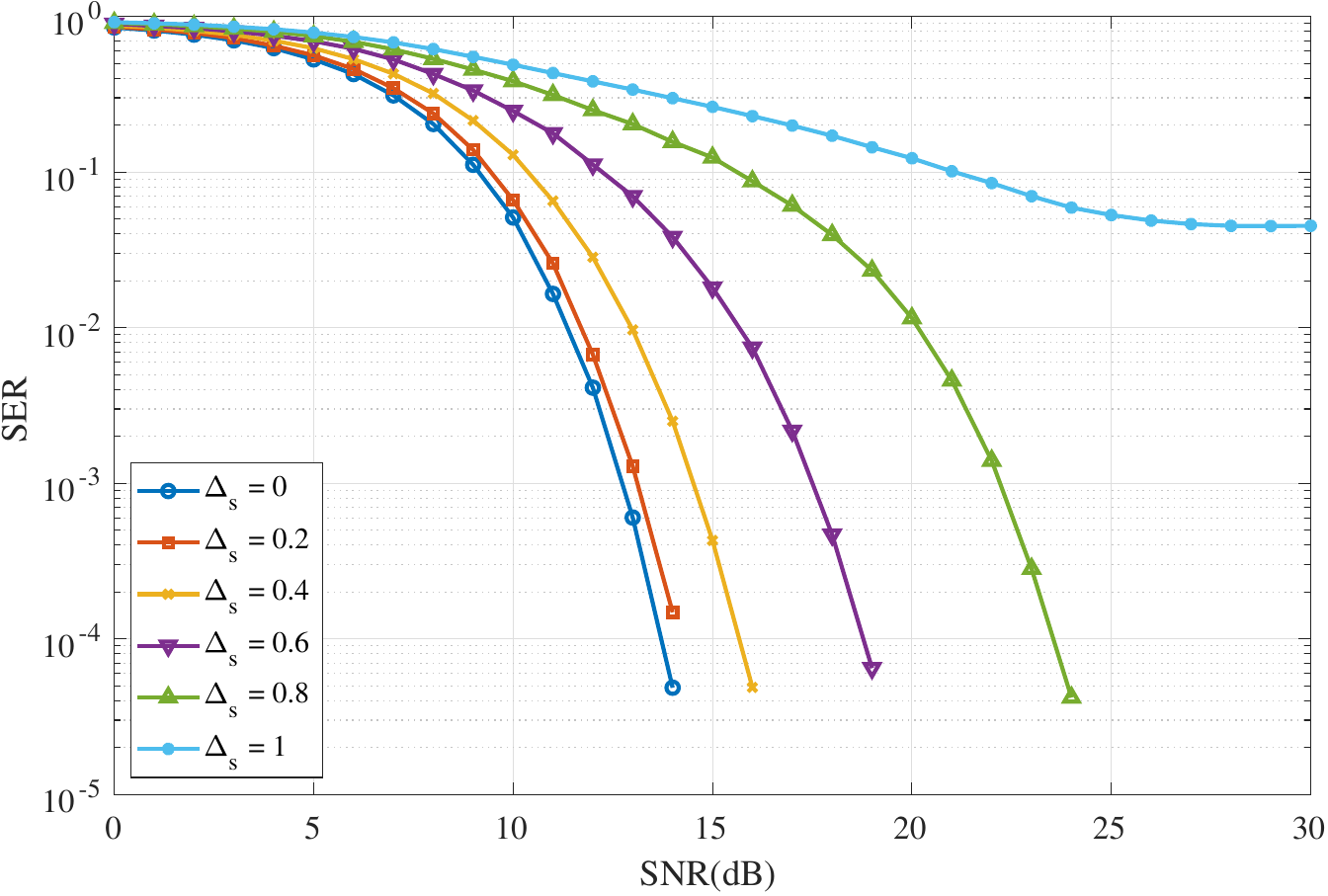}
        \caption{$\mathrm{SF}={5}$ and raised-cosine chip waveform}
        \label{fig:6}
    \end{subfigure}
    \caption{SER performance of quasisynchronous LoRa communications over AWGN channels; $\mathrm{SF}={4,5}$}
    \label{fig:allfigures45}
    \vspace{-4mm}
\end{figure*}
\section{Simulation and Numerical Results}
In this section, we present a detailed analysis of SER performance of synchronous and quasisynhronous LoRa communication over AWGN channels through Monte-Carlo simulations conducted in Matlab environment. We evaluate the performance across various spreading factors ($\mathrm{SF}={4,5,6,7}$). Further, we consider two distinct chip-limited chip waveforms, namely rectangular and raised-cosine waveforms. The simulation results are obtained by varying the maximum normalized quasisynchronization error $\Delta_s$ among the following values: $0, 0.2, 0.4, 0.6, 0.8$, and $1$.

\begin{figure*}[t]
    \centering
    \begin{subfigure}{0.45\textwidth}
        \includegraphics[width=\linewidth]{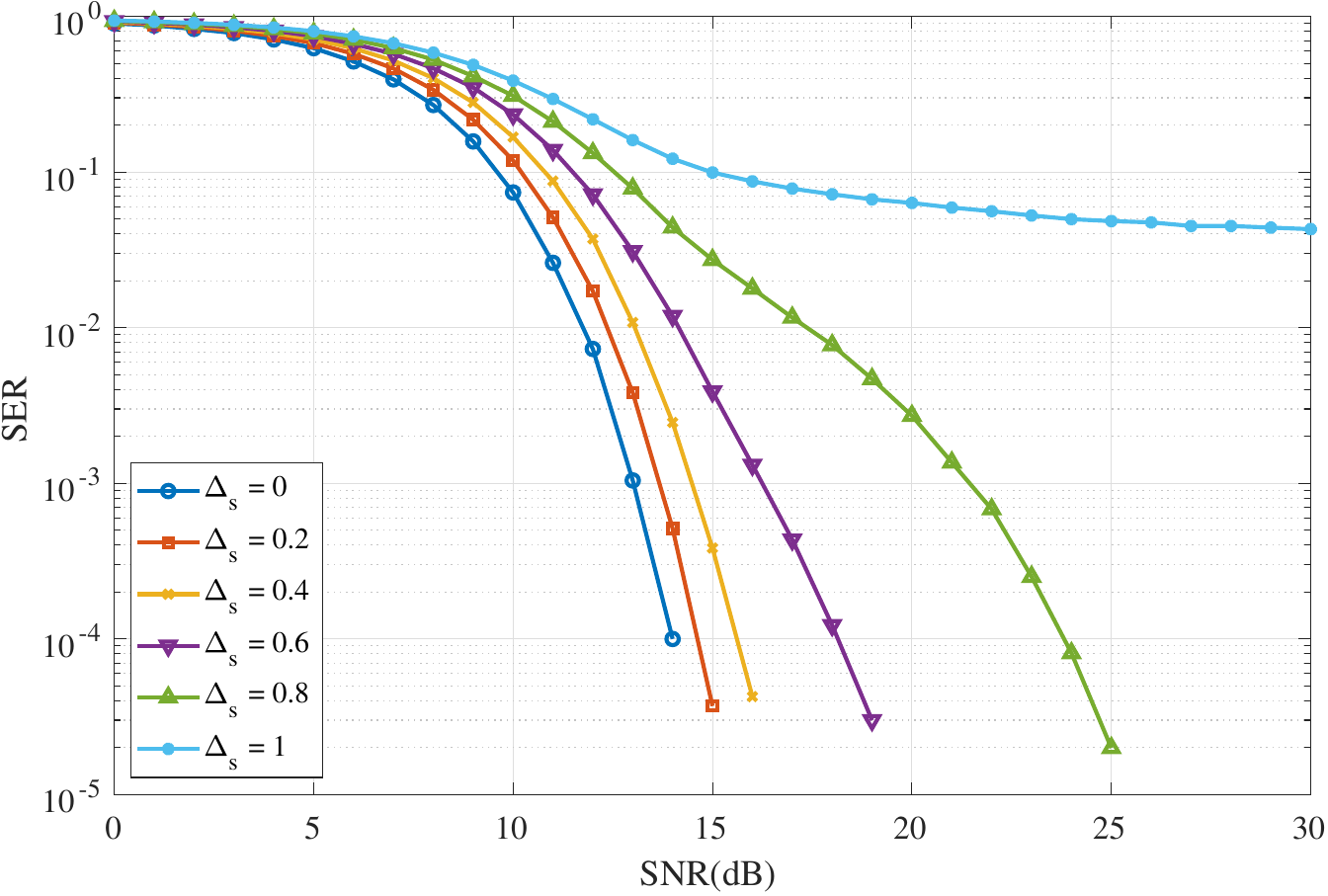}
        \caption{$\mathrm{SF}={6}$ and rectangular chip waveform}
        \label{fig:7}
    \end{subfigure}
    \hfill
    \begin{subfigure}{0.45\textwidth}
        \includegraphics[width=\linewidth]{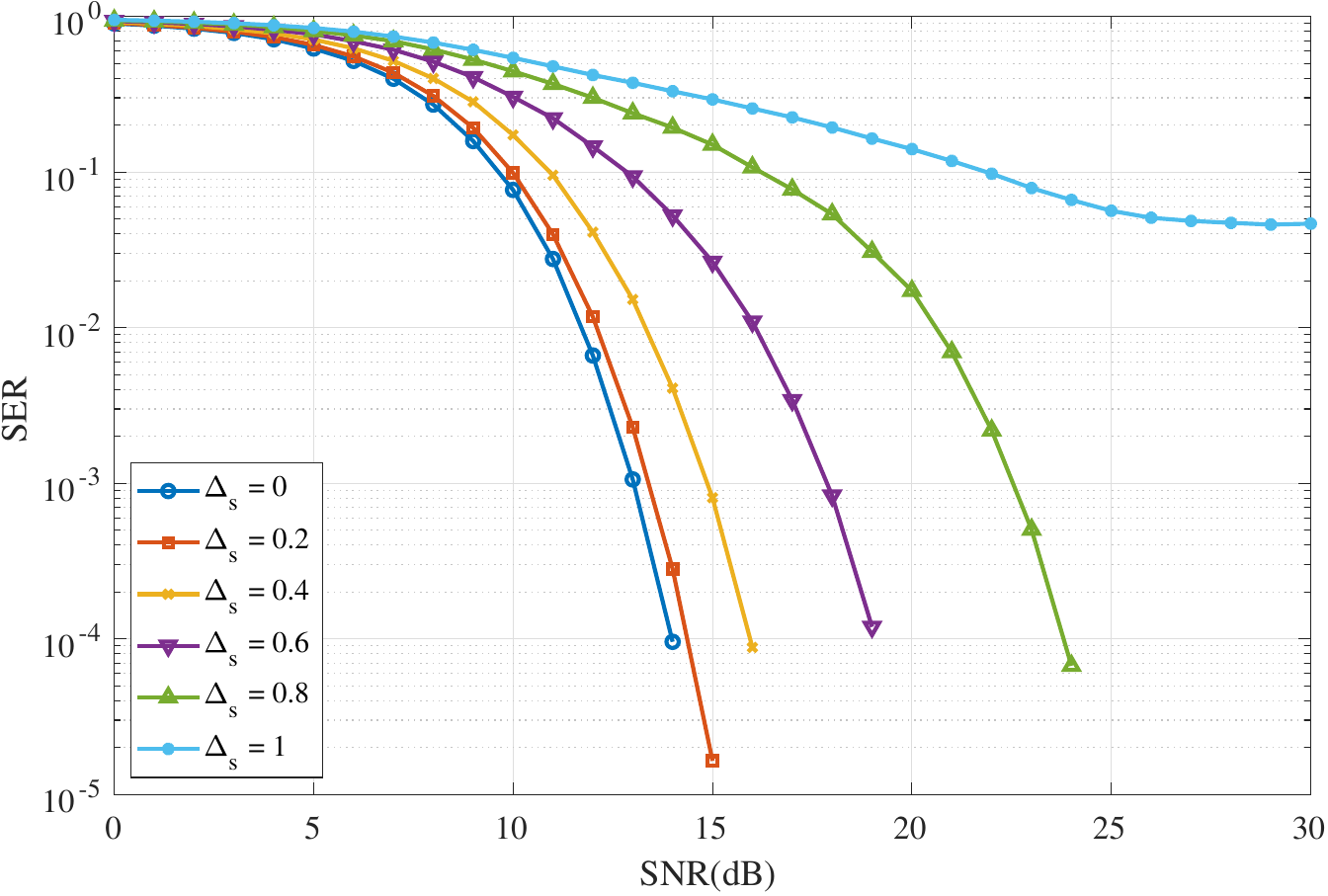}
        \caption{$\mathrm{SF}={6}$ and raised-cosine chip waveform}
        \label{fig:8}
    \end{subfigure}
    
    \vspace{6pt} 
    
    \begin{subfigure}{0.45\textwidth}
        \includegraphics[width=\linewidth]{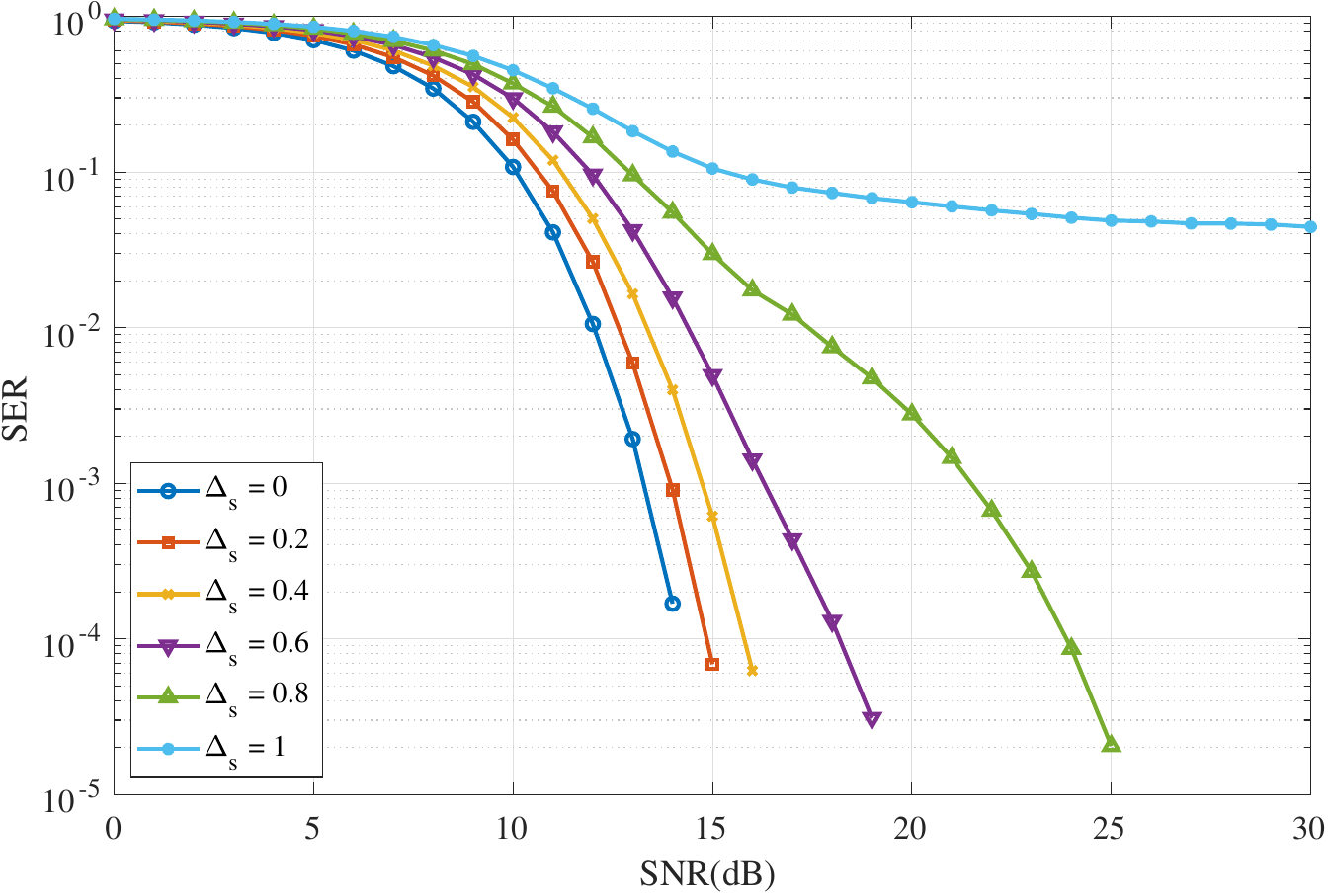}
        \caption{$\mathrm{SF}={7}$ and rectangular chip waveform}
        \label{fig:9}
    \end{subfigure}
    \hfill
    \begin{subfigure}{0.45\textwidth}
        \includegraphics[width=\linewidth]{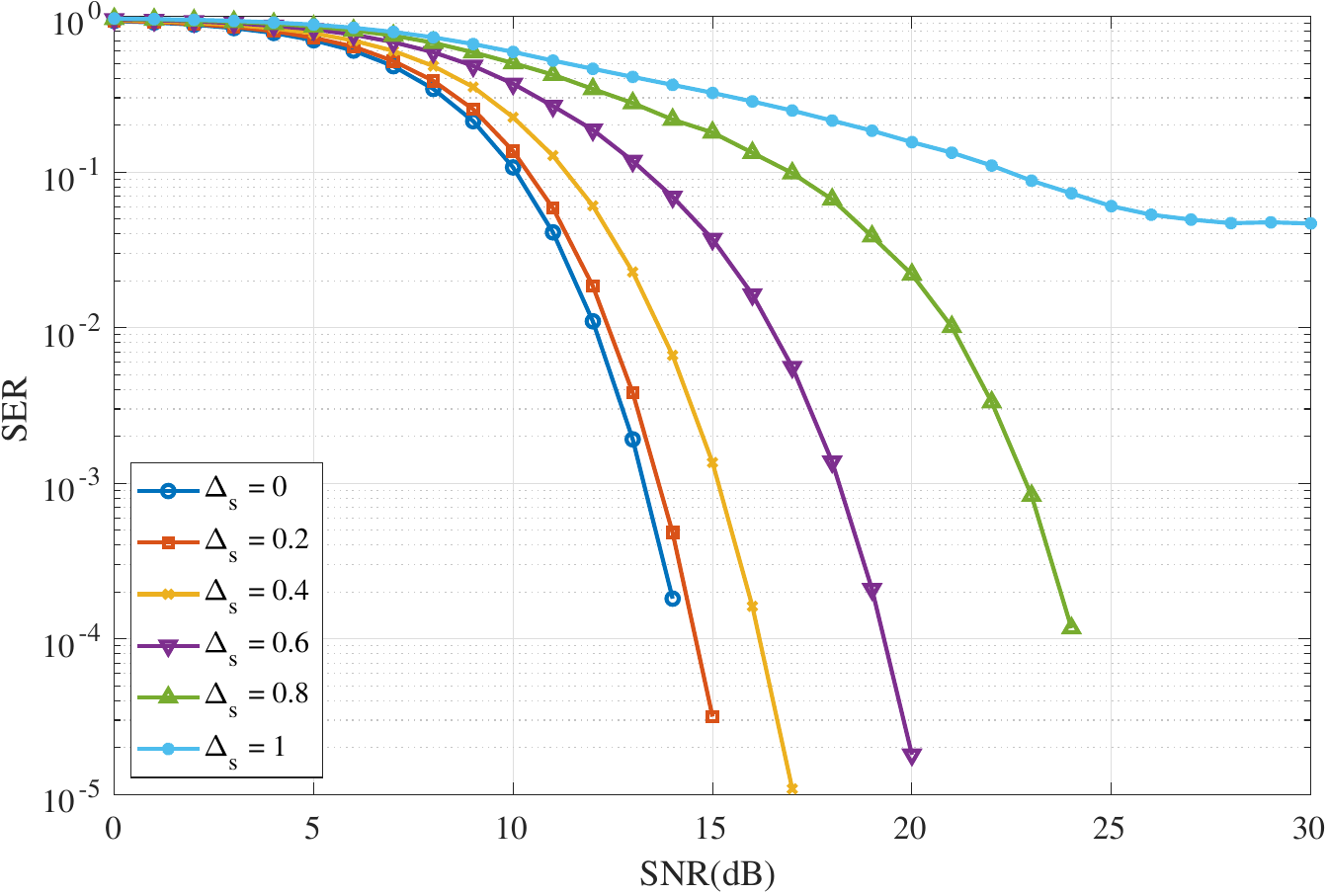}
        \caption{$\mathrm{SF}={7}$ and raised-cosine chip waveform}
        \label{fig:10}
    \end{subfigure}
    \caption{SER performance of quasisynchronous LoRa communications over AWGN channels; $\mathrm{SF}={6,7}$}
    \label{fig:allfigures67}
    \vspace{-4mm}
\end{figure*}

The simulation results are depicted in Fig. \ref{fig:allfigures45} and Fig. \ref{fig:allfigures67}. It is observed that the required Signal-to-Noise Ratio (SNR) to achieve a target SER is nearly similar for each spreading factor. This is due to the fact that the LoRa waveforms have unit energy for each symbol. For example, to achieve an SER of $10^{-3}$ for $\mathrm{SF}=6$ and $\mathrm{SF}=5$ using the rectangular chip waveform, the required SNR is approximately 14 dB for $\Delta_s=0.4$. The impact of $\Delta_s$ is evident in Fig. \ref{fig:allfigures45} and Fig. \ref{fig:allfigures67}, where it can be seen that as $\Delta_s$ increases, the SER performance deteriorates. Although the SER performance with $\Delta_s=0$ is better than with $\Delta_s>0$ for all spreading factors and chip waveforms, reliable communication can still be achieved with $\Delta_s<1$. However, an error floor is observed for $\Delta_s=1$ in Fig. \ref{fig:allfigures45} and Fig. \ref{fig:allfigures67}. Even at high SNRs, the SER for $\Delta_s=1$ remains greater than $10^{-2}$ for all spreading factor values and chip waveforms. To achieve an SER lower than $10^{-4}$, it is necessary to have $\Delta_s<1$ for all spreading factors and chip waveforms. In addition to the quasi-synchronous performance, the impact of the chip waveform exhibits different characteristics. For all spreading factors, the raised-cosine chip waveform outperforms the rectangular chip waveform when $\Delta_s=0.2$. However, for higher values of $\Delta_s$, the performance of the rectangular chip waveform becomes better than that of the raised-cosine chip waveform. For instance, Fig. \ref{fig:3} and Fig. \ref{fig:4} illustrate that for $\mathrm{SF}=4$, to achieve an SER of $10^{-3}$, the required SNR is approximately 11 dB with the rectangular chip waveform and 13 dB with the raised-cosine chip waveform. Consequently, it can be inferred that as $\Delta_s$ decreases, the SER performance of the raised-cosine chip waveform surpasses that of the rectangular chip waveform.

Based on these observations, we can conclude that the selection of spreading factor and synchronization accuracy, represented by $\Delta_s$, significantly influence the SER performance of LoRa communication systems. Keeping a lower $\Delta_s$ value can result in improved SER performance, even when perfect synchronization is unattainable. This information holds great value for system designers and engineers aiming to optimize LoRa communication for specific applications and environments.

\section{Conclusion}
The paper focuses on implementing quasisynchronous communication to LoRa systems for ground-to-LEO satellite communication, addressing the increased need for IoT applications and worldwide connectivity. We elaborated different chip waveform performances, including rectangular and raised-cosine waveforms, and we presented the SER performance of quasisynchronous LoRa communication for different spreading factors in AWGN channel. When perfect synchronization is not achievable, communication reliability is ensured with an acceptable quasisynchronization error

\bibliographystyle{IEEEtran}
\bibliography{conference_101719}

\end{document}